\documentstyle[12pt]{article} 
\normalbaselines

\begin{document}

\newtheorem{lema}{Lemma}[section]
\newtheorem{prop}[lema]{Proposition}
\newtheorem{teor}[lema]{Theorem}
\newtheorem{coro}[lema]{Corollary}

\font\ddpp=msbm10 scaled \magstep 1 
\def\R{\hbox{\ddpp R}}    
\def\C{\hbox{\ddpp C}}     
\def\L{\hbox{\ddpp L}}    
\def\N{\hbox{\ddpp N}}    
\def\Z{\hbox{\ddpp Z}}    


\bibliographystyle{unsrt} 

\vbox {\vspace{6mm}} 

\begin{center}
{\large \bf SOME REMARKS ON CAUSALITY THEORY AND VARIATIONAL METHODS IN LORENTZIAN MANIFOLDS}  \\[2mm]
Miguel S\'anchez\\ 
{\it Departamento de Geometr\'{\i}a y Topolog\'{\i}a\\ Facultad de Ciencias, 
Universidad de Granada, 18071-Granada, Spain.}\\ 
e-mail address: sanchezm@goliat.ugr.es\\[5mm]
\end{center}

\section{Introduction.} 

A geodesic on a semi-Riemannian manifold is a extremal curve of the 
energy functional. In the Lorentzian case, causal geodesics (see definitions 
in Section 2) are also curves which maximizes locally the time-separation, in a similar way as the geodesics of a Riemannian manifold minimize locally the associated distance. So, when certain problems about causal geodesics are studied, the next two points of view appear naturally. The first one, which we will denote CT (Causality Theory) exploits the properties of the time-separation. For example, the next results have been obtained by CT: 
\newcounter{causal}
\begin{list}{CT--\arabic{causal}}
{\usecounter{causal}}
\item In a globally hyperbolic spacetime there exists a causal geodesic 
joining each couple of causally related points \cite{Av}, \cite{Se}.
\item The existence of causal closed geodesics in certain compact 
Lorentzian manifolds \cite{Ga1}, \cite{Ga2}, \cite{Ti}. 
\end{list}
In CT the causal geodesics are characterized by a specific property of them, with a clear interpretation. Under the second point of view or VM (Variational Methods), causal geodesics are treated as particular cases of the extremal points of the energy functional and, after, their specific  properties are taken into account. The next two topics have been studied by VM:
\newcounter{vm}
\begin{list}{VM--\arabic{vm}}
{\usecounter{vm}}
\item The existence and multiplicity of timelike geodesics joining two given points in certain spacetimes, as static, stationary and splitting type, see for example, \cite{Ma-li}. 
\item The existence of causal periodic trajectories \cite{BF}, \cite{BFG-ihp}, \cite{Ca}, \cite{Ma-ed}, \cite{Sa-ed}.
\end{list}
In fact, the machinery of VM can be also applied to study non-causal geodesics and, so, the geodesic connectedeness of some spacetimes have been studied with it. Moreover, VM consider all the critical points of the energy (not only the maxima) and, so, it seems especially appropriate to study multiplicity results. In \cite{Sa-ca} a relation between VM--2 and CT--2 is studied. Our  purpose here is to study the relation between VM--1 and CT--1.

This paper is organized as follows. In Section 2 some notation and generalities are introduced, and we focus our attention in a general kind of product spacetimes ${\cal M} = \R{}\times {\cal M}_{0}$, usually studied by VM. In Section 3 some natural conditions for the global hyperbolicity of such spacetimes are obtained, Proposition 3.1-- Proposition 3.5. In Section 4 a function representing the ``universal time" to travel from a point $z\in {\cal M}$ to a ``static trajectory" $\{ (t,x)/t\in \R{} \} \subset {\cal M}$ is introduced. When this function remains finite, some results on geodesic connectedeness (Corollary 4.4), multiplicity of timelike geodesics (Proposition 4.5) or non-existence of causal geodesics (Proposition 4.7) are easily derived by using CT. In the last section, some open questions are stated, comparing the previous results with those obtained by VM.

\section{Generalities about spacetimes and notation.}

In this Section some well-known elemental properties about causality  in spacetimes are reviewed; we also introduce the notation and define the spacetimes to be studied. Basic references as \cite{BE}, \cite{On} will be followed.

A Lorentzian manifold $({\cal M},g)$ is a manifold $\cal{M}$ endowed with a non-degenerate metric $g$ of index 1, $(-,+,\dots,+)$. A Lorentzian manifold is a particular case of semi-Riemannian manifold, that is, a manifold with a metric which is assumed to be just non-degenerate and with constant index. In what follows, all the manifolds are connected, and all the objects are smoothly differentiable, except when something else is explicitly said. A tangent vector $v \in T{\cal M}$ is timelike (resp. lightlike; causal; spacelike) if $g(v,v)<0$ (resp. $g(v,v)=0$, and $v \neq 0$; $g(v,v) \leq 0$, and $v \neq 0$; $g(v,v)>0$ or $v=0$); note that, following \cite{On}, the vector 0 is defined as spacelike. 

We will also assume that $({\cal M},g)$ is time-orientable, and that a time-orientation (continuous choice of causal cones, which are called future cones) has been chosen; then, we will call spacetime to this time-oriented Lorentzian manifold. Given $z_{1}, z_{2} \in {\cal M}$, we will denote, as usual, $z_{1}<<z_{2}$ (resp. $z_{1}<z_{2}$; $z_{1} \leq z_{2}$) if there exist a future-pointing timelike (resp. causal; causal or constant) curve from $z_{1}$ to $z_{2}$; in this case, both points are said to be chronologically related (resp. causally related). The relations $<<, < $ and $ \leq$ follow natural transitive relations; for example, $z_{1}<<z_{2}$ and $z_{2} \leq z_{3}$ imply $z_{1}<<z_{3}$. 
Given any subset $A \subset {\cal M}$, its chronological and causal futures are defined, respectively, as:  
\[
I^{+}(A) = \{ z \in {\cal M}: \exists z'\in A,\ {\rm such \ that}\  z'<<z\},\   
J^{+}(A) = \{ z \in {\cal M}: \exists z'\in A,\ {\rm such \ that}\  z'\leq z\}
\] 
(the past subsets $I^{-}(A),\  J^{-}(A)$ are defined dually). 

The length of a smooth curve $\gamma: [a,b]\rightarrow {\cal M}$ is the integral: 
\[
L(\gamma) = \int_{a}^{b}  \mid g(\gamma',\gamma')\mid ^{1/2}
\]
The time-separation or Lorentzian distance function $\Xi : {\cal M \times M} \rightarrow [0,\infty ]$ is defined as:
\[
\Xi (z_{1},z_{2})= Sup\{ L(\gamma): \gamma {\rm  \ is \  causal \ and  \ future-directed, \ with} \  \gamma (a)= z_{1}, \gamma (b) = z_{2} \}.
\]
If $z_{2} \not\in I^{+}(z_{1})$, then its time-separation is 0. The time-separation may be infinite; in fact, if $\Xi \equiv \infty$ the spacetime is called totally vicious.   

A Cauchy surface is  a subset $S$ (necessarily a closed and connected topological hypersurface) which is met exactly once by any inextendible causal curve. A spacetime containing such a surface is called globally hyperbolic (this definition has several classical equivalences, which is not relevant to discuss here), and it is homeomorphic to $\R{} xS$; moreover, 
all its Cauchy surfaces are homeomorphic \cite{Ge}. In a globally hyperbolic spacetime the time-separation is finite (although not necessarily bounded) and continuous, and the subsets $J^{+}(z_{1}) \cap J^{-}(z_{2})$ are compact. Even more, in a suitable sense the set of continuous non-spacelike curves joining $z_{1}$ and $z_{2}$ is compact, which is the key to yield CT--1 in Section 1. Note that the causal relations and Cauchy surfaces are invariant under conformal transformations of the metric, even though $\Xi$ is not.

Consider an open subset $U\subset {\cal M}$ as a spacetime. Of course, its time-separation $\Xi _{U}$ may be different to the restriction $\Xi \mid_{U}$ of the time-separation of $M$ to $U$; in general, $\Xi _{U} \leq \Xi \mid_{U}$. In fact, for each $p\in {\cal M}$ one can find a neighborhood $U \ni p$ with a single causal behaviour as spacetime: if $z_{1}, z_{2} \in U$ then $z_{1} < z_{2}$ in $U$ if and only if there exist a causal geodesic $\gamma$ in $U$ joining them, which is unique up to reparametrizations and satisfies $L(\gamma)= \Xi _{U}(z_{1},z_{2})$.  In this sense, causal geodesics maximizes locally the time-separation. 
 
\vspace{3mm}

\noindent In what follows we will consider the next kind of spacetime: ${\cal M}$ is the product of the real numbers by another manifold ${\cal M = \R{} \times M}_{0}$, and the metric $g$ at each $z = (t,x) \in {\cal M}$ is:
\[
g((\tau,\xi),(\tau,\xi)) = - \beta (z) \tau ^{2} + 2 <\delta (z),\xi > \tau + <\alpha _{z}(\xi ),\xi >, \; \forall (\tau,\xi )\in T_{z}{\cal M} \equiv \R{} \times T_{x}{\cal M}_{0}, 
\]
where $\beta$ is a positive function on ${\cal M}$, $<\cdot ,\cdot >$ denotes a Riemannian metric on ${\cal M}_{0}$ (with associated norm $\parallel \cdot \parallel$ and distance {\it dist$(\cdot ,\cdot )$}), $\alpha _{z}$ is a symmetric positive operator on 
$T_{x}{\cal M}_{0}$ and $\delta (z) \in T_{x}{\cal M}_{0}$, all varying smoothly with $z$. The maximum (resp. minimum) eigenvalue of $\alpha _{z}$ will be denoted $\lambda _{max}(z)$ (resp. $\lambda _{min}(z)$); note that the eigenvalues of $\alpha _{z}$ vary (a priori just) continuously with $z$.

Redefining, if necessary, $\delta$ and $\alpha$, there is no loss of generality in assuming that $<\cdot ,\cdot >$ is complete (see the Example A in Section 3), as we will do from now on. It is a single exercise to check that, under our assumptions, $g$ is always Lorentzian. We will chose the time-orientation given by the natural vector field $\partial _{t}$, and the function $t$ (projection of ${\cal M}$ onto $\R{}$) can be seen as a sort of {\it universal time}. 

As particular cases of interest, when $g$ is indepent of the variable $t$, the spacetime is called stationary. In this case, we will write $g_{R} = \ <\alpha(\cdot),\cdot>$ (which may be incomplete). If, moreover, $\delta \equiv 0$ then it is called static; that is, in this case (with natural identifications) $g= -\beta dt^{2} + g_{R}$, where $\beta$ is a function on ${\cal M}_{0}$. 

Note that many physically important spacetimes are isometric to products as above; in fact, every globally hyperbolic spacetime is expected to be isometric to one of them. As well-known examples, we have: Robertson-Walker (and its generalized versions in \cite{ARS}), Schwarzschild (as the outer as the inner region), Reissner-Nordstr\"{o}m  (each one of the three regions) and Kerr spacetimes (for a more specific study of spacetimes as the second region of  Reissner-Nordstr\"{o}m one, see \cite{Gi}, \cite{GM-man}, \cite{Sa-grg}).

\section{Sufficient conditions for global hyperbolicity.}

Next, our aim will be to study when each slice ${\cal M}_{t} = \{ t\} \times {\cal M}_{0} \subset {\cal M}$ is a Cauchy surface. The following single example may be useful. 

\vspace{3mm}

\noindent {\it Example A.} Consider an incomplete Riemannian manifold $({\cal M}_{0},g_{R})$, and the Lorentzian product 
$(\R{} \times {\cal M}_{0}, -dt^{2}+g_{R})$.     
Clearly, for each slice ${\cal M}_{t_{0}} = \{ t_{0} \} \times {\cal M}_{0}$ one can find an inextendible causal curve $\gamma(s) = (t(s),x(s)) $ with $t(s)<t_{0}$ for all $s$. So, $\gamma$ does not crosses it and, thus, ${\cal M}_{t_{0}}$ is not a Cauchy surface, Fig. 1 (in fact, this spacetime is not globally hyperbolic, see Proposition \ref{prop 3.5}). Now, rewrite this example with a complete metric for ${\cal M}_{0}$ as follows. Take a factor $\Omega >0$ such that  $<\cdot,\cdot> = \Omega \cdot g_{R}$ is complete (recall that every Riemannian metric is conformal to a complete one, \cite{NO}), and put $\alpha = {\rm Id}/\Omega $. Now, consider $\gamma$ parametrized by the $t$ variable, $\gamma(t) = (t,x(t)) $. As $\gamma$ is inextendible, the range of $x(t)$ is not imprisoned in any compact subset and, as $<\cdot,\cdot>$ is complete, its length is $\infty$. Thus, 
$\parallel x'(t) \parallel $ must diverge  enough fast (Fig. 2). $\Box$

\vspace{3mm}

\noindent Now, consider our general product spacetime $({\cal M},g)$ and any causal curve $\gamma : ]a,b[ \rightarrow {\cal M}$, $\gamma(t) = (t,x(t))$. Following the Example A, we will impose to $\parallel x'(t) \parallel $ ``not to diverge too fast". As $g(\gamma ',\gamma ') \leq 0$ we have: 
\begin{equation}
\label{3.1}
-\beta + 2<\delta,x'> + \lambda_{min} \parallel x' \parallel ^{2} \leq 0 . 
\end{equation}
at each $(t,x(t))$; moreover, by using Cauchy-Schwarz inequality:
\begin{equation}
\label{3.2}
\lambda_{min} \parallel x' \parallel ^{2} - 2 \parallel \delta \parallel \parallel x' \parallel - \beta \leq 0 . 
\end{equation}   
Thus, $\parallel x'\parallel$ can not be greater than the bigger root of the equality obtained from (\ref{3.2}), that is: 
\begin{equation}
\label{3.3}
\parallel x' \parallel \; \leq \; (\parallel \delta \parallel + (\lambda _{min} \beta + 
\parallel \delta \parallel  ^{2})^{1/2})/\lambda _{min}. 
\end{equation}   
This single fact, yields the next result. 

\begin{prop}
\label{prop 3.1}
For each positive integer $n$, put ${\cal M}[n] = [-n,n]\times {\cal M}_{0} \subset {\cal M}$. If there exists a smooth function $F_{n}$ on ${\cal M}_{0}$ for each $n$ such that: 

(i) the next inequality holds for all $(t,x) \in {\cal M}[n])$: 
\begin {equation}
\label{fn}
\frac{\parallel \delta \parallel + (\lambda_{min} \beta + \parallel \delta \parallel ^{2})^{1/2}}{\lambda _{min}}(t,x) \leq F_{n}(x), 
\end{equation}

(ii) the metric $<\cdot ,\cdot >_{n} = <\cdot , \cdot>/F_{n}^{2} \, $ on ${\cal M}_{0}$ is complete, 

\noindent then each slice ${\cal M}_{t} = \{ t\} \times {\cal M}_{0}$ is  a Cauchy surface.
\end{prop}

\pagebreak

{\it Proof.} Otherwise, there exists an inextendible causal curve $\gamma: ]a,b[ \rightarrow {\cal M}$, $\gamma(t) = (t,x(t))$ with $a$ or $b$ finite. Assume $b< \infty $ (the other case is analogous), and choose an integer $n$ and a scalar $a'<b$ such that $-n< b \leq n$, Max$\{ -n,a \} < a'$. Then, by (\ref{3.3}) and  (\ref{fn}): 
\[
\parallel x' \parallel \leq F_{n}(x(t)), \; \forall t\in [a',b[, 
\]
that is, $<x',x'>_{n} \, \leq 1$. Thus, this inextendible curve has finite length for $<\cdot , \cdot>_{n}$, in contradiction with {\it (ii)}. $\Box$

\vspace{3mm}

\noindent It is easy to impose reasonable sufficient conditions for the existence of such $F_{n}$ and, thus, for the global hyperbolicity of the spacetime. Fix a point $x_{0} \in {\cal M}_{0}$ and denote by $d_{0}: {\cal M}_{0} \rightarrow \R{}$ the $<\cdot , \cdot>$--distance function to $x_{0}$. Then, put:

\[
M_{n} = {\rm Sup}\{ \frac{\parallel \delta \parallel}{\lambda_{min} \, d_{0}}(t,x), \sqrt{\frac{\beta}{\lambda_{min} \, d_{0}^{2}}}(t,x): (t,x)\in {\cal M}[n], d_{0}(x)>1 \} \in [0,\infty] 
\]

\noindent (If the diameter of ${\cal M}_{0}$ is not greater than 1, put 
$M_{n}=0$ for all $n$.) 
Note that the finiteness of $M_{n}$ is independent of the chosen point $x_{0}$. 

\begin{prop}
\label{prop 3.2}
If the constants $M_{n}$ above are all finite, then a set of functions $F_{n}, \, n\in \N{}$, as in Proposition \ref{prop 3.1} can be found. 
\end{prop} 

{\it Proof.} Choose functions $F_{n}= (1+ \sqrt{2}) M_{n} \cdot d_{0}$ out a suitable compact subset to obtain (\ref{fn}). The completeness of the metrics $<\cdot , \cdot >_{n} $ can be proven by checking that the corresponding length of any diverging curve (say, starting at $x_{0}$,  with $<\cdot , \cdot >$-speed equal to 1) is $\infty .$$ \Box$
  
\vspace{3mm}

\noindent {\it Remark.} Proposition \ref{prop 3.2} impose to $\parallel \delta \parallel / \lambda_{min}$ and $\sqrt{\beta / \lambda_{min}}$ not to diverge in the $x$ variable faster than $d_{0}(x)$; nevertheless, as $F_{n}$ may change with $n$, it does not matter if they diverge fastly with $t$. 

\vspace{3mm}

\noindent As  straightforward consequences, we have:

\begin{coro}
\label{coro 3.3}
If ${\cal M}_{0}$ is compact then $({\cal M},g)$ is globally hyperbolic.
\end{coro}

\begin{coro}
\label{coro 3.4}
Assume that $({\cal M},g)$ is stationary. If 
\begin{equation}
\label{3.4a}
{\rm Sup}\{ \frac{\parallel \delta \parallel}{\lambda_{min} \, d_{0}}(x), \sqrt{\frac{\beta}{\lambda_{min} \, d_{0}^{2}}}(x): x\in {\cal M}_{0}, d_{0}(x)>1 \}
\end{equation}

\noindent is finite, then the spacetime is globally hyperbolic.

In particular, it occurs if the induced metric $g_{R} = \ <\alpha(\cdot), \cdot >$ is complete and there exist constants $a, b, c, d$ such that, for $x$ out a compact subset:
\begin{equation}
\label{3.4b}
\parallel \delta (x) \parallel < a d_{0}(x) + b , \; \; {\rm and} \; \; 
 \sqrt{\beta }(x) < c d_{0}(x) + d . 
\end{equation}

\end{coro}

\vspace{3mm}

\noindent The static case is simpler and the natural sufficient condition becomes also necessary.

\begin{prop}
\label{prop 3.5}
The slices of a static spacetime $({\cal M}=\R{}\times {\cal M}_{0}, g=-\beta dt^{2} + g_{R})$ are  Cauchy surfaces if and only if the metric $g_{R}/\beta$ is complete. 
\end{prop}

{\it Proof.} Note that $g$ is conformal to the product metric  $-dt^{2} + g_{R}/\beta$, which is globally hyperbolic if and only if $g_{R}/\beta$ is complete, \cite[Theorem 2.53]{BE}. $\Box$ 

\vspace{3mm}

\noindent {\it Remark.} (1) Note that, if $g_{R}$ in Proposition \ref{prop 3.5} is complete, then the condition on $\beta$ in (\ref{3.4b}) implies the completeness of  $g_{R}/\beta$. 

(2) Proposition \ref{prop 3.5} can be extended with the same proof replacing $\R{}$ by an open interval, that is, ${\cal M}= ]a,b[ \times {\cal M}_{0}, -\infty \leq a < b \leq \infty $. The extensions of the Propositions \ref{prop 3.1} and \ref{prop 3.2} to this case are also straightforward and, as a consequence, the Corollaries \ref{coro 3.3} and \ref{coro 3.4} still hold then. A more general version of Proposition \ref{prop 3.5} can be seen in \cite[Proposition 3.7]{Be-Po}.

\vspace{3mm}

\noindent {\it Examples (B.1).} The two dimensional anti-de Sitter spacetime, defined by  ${\cal M}_{0} = ]-\pi /2,\pi /2[$, $g=-dt^{2}/cos^{2}(x) + dx^{2}/cos^{2}(x)$ , is static, complete (the metric $g_{R}=dx^{2}/cos^{2}(x)$ is complete too), and non geodesically connected.  It is clear from Proposition \ref{prop 3.5} that it is not globally hyperbolic. Changing the function $\beta \equiv cos^{-2}$ by a new positive function $\beta^{*}$ satisfying 
\[
\int_{0}^{\pi /2}  (\beta^{*} \, cos^{2})^{-1}(x)\, dx = \int^{0}_{-\pi /2} (\beta^{*} \,  cos^{2})^{-1}(x)\, dx  = \infty 
\]
\noindent a globally hyperbolic spacetime is obtained. (The geodesic connectedeness of metrics extending this example is studied in detail in \cite{Sa-grg} and the geodesic completeness in \cite{RS-gd}.) 

{\it (B.2).} Consider a {\it warped} metric $g=-dt^{2} + f(t)^{2} g_{R}$, where $f$ is a positive function on (an interval of) $\R{}$. If the Riemannian metric $g_{R}$ on ${\cal M}_{0}$  is complete, then the constants $M_{n}$ in Proposition \ref{prop 3.2} are trivially finite, and the spacetime becomes globally hyperbolic. Otherwise, if $g_{R}$ is incomplete, the spacetime is not globally hyperbolic. One can check it from the {\it Remark} (2) above,  putting $g^{*}= g/f^{2}$ and $ ds=dt/f$. Then $g^{*} = ds^{2} + g_{R}$ and, so, neither $g^{*}$ nor the conformal metric $g$ are globally hyperbolic (see \cite[Theorem 2.55]{BE} for a more general result).

\section{Joining a point and a static trajectory.}

Under the sufficient conditions for global hyperbolicity in Section 3, 
one can claim that two points can be joined by a causal geodesic if and only if they are causally related. Now, we are going to study when this causal relation holds. As an application, some multiplicity results for the existence of timelike geodesics between points 
$(t_{1}, x_{1}),(t_{2},x_{2})$  with $\mid t_{1}-t_{2} \mid $ large will be yielded. Throughout all this Section we will assume that each slice ${\cal M}_{t}$ is a Cauchy surface. 

\vspace{3mm}

\noindent Consider the {\it future arrival time} function $T_{0}:  {\cal M} \times  {\cal M}_{0} \rightarrow [0,\infty]$, 
\begin{equation}
\label{t0}
T_{0}((t_{1},x_{1}),x_{2}) = 
{\rm Inf}\,\{ t-t_{1}: (t_{1}, x_{1})\leq (t,x_{2}), t\in \R{} )\},  
\end{equation}
If $(t_{1}, x_{1}) \not\leq (t,x_{2})$ for all $t$ then the value of $T_{0}$ is $\infty$. Intuitively, this function assigns the infimum universal time to travel between $(t_{1}, x_{1})$ and the {\it static trajectory} $R_{x_{2}}=\{ (t,x_{2}): t\in \R{} \} $. Dually, the {\it past arrival time} function $T_{0}^{*}$ is defined by: 
\begin{equation}
\label{t0'}
T_{0}^{*}((t_{1},x_{1}),x_{2}) = 
{\rm Inf}\,\{ t_{1}-t: (t, x_{2})\leq (t_{1},x_{1}), t\in \R{})\} ,  
\end{equation}
In a natural way, $T_{0}^{*} $ is the future arrival time function for the spacetime obtained reversing the time-orientation. 
The properties of $T_{0}$ have been studied in \cite{Sa-ca}, obtaining in particular: 

\begin{lema}
\label{lema 4.1}
(i) The relation $ (t_{1}, x_{1})\leq (t_{2},x_{2})$ holds if and only if $T_{0}((t_{1}, x_{1}),x_{2}) \leq t_{2}-t_{1}$, with equality if and only $(t_{2}, x_{2})\in J^{+}(t_{1},x_{1})$ \verb+\+ $I^{+}(t_{1},x_{1})$. 

(ii) $T_{0}$ is a continuous function in both variables. 
\end{lema}

\noindent Next, we are going to study when $T_{0}$ is finite. As a first property, one has:

\begin{lema}
\label{lema 4.2}
If $T_{0}((t_{1},x_{1}),x_{2}) < \infty$ then $T_{0}((t,x_{1}),x_{2}) < \infty$ for all $t\leq t_{1}$.
\end{lema}

{\it Proof.} By hypothesis there exists $t_{2}>t_{1}$ such that $(t_{1},x_{1})\leq (t_{2},x_{2})$. As $(t,x_{1})\leq (t_{1},x_{1})$ for any $t\leq t_{1}$, the result follows from the transitivity of the causal relations. $\Box$

\vspace{3mm}

\noindent The next result yields a natural sufficient condition for the finiteness of $T_{0}$.

\begin{prop}
\label{prop 4.3}
If there exists a continuous function 
$K: {\cal M}_{0} \rightarrow ]0,\infty[ $ such that:
\begin{equation}
\label{bound}
\frac{\parallel \delta \parallel}{\beta}(t,x) \leq K(x)t, \; \; \; 
\sqrt{\frac{\lambda_{max}}{\beta}}(t,x) \leq K(x)t
\end{equation}

\noindent $\forall x\in {\cal M}_{0}, \, \forall t>1$, then the function $T_{0}$ does not reach the value $\infty$.
\end{prop}

{\it Proof.} Take $(t_{1},x_{1})\in {\cal M}, x_{2}\in {\cal M}_{0}$, and let $x:[0,b]\rightarrow {\cal M}_{0}$ be a curve joining $x_{1}$ and $x_{2}$ with $<x',x'> \equiv 1$. Consider the equation in $t$ obtained by imposing that $(t,x)$ is a future-directed lightlike curve joining $(t_{1},x_{1})$ and the static trajectory $R_{x_{2}}$, that is:
\begin{equation}
\label{4.3a}
t'(s) = \frac{-<\delta ,x'> + \sqrt{<\delta,x'>^{2} + \beta <\alpha (x'),x'>}}{\beta}(t(s),x(s))
\end{equation}
with initial condition $t(0)=t_{1}$. As $x$ is fixed, call $f(t,s)$ to the right hand side of (\ref{4.3a}). It is known from elemental theory of equations, that there exist a solution to (\ref{4.3a}) defined in all $[0,b]$ if there is a positive number $k$ such that: 
\begin{equation}
\label{4.3 cielos}
f(t,s)-f(t_{1},s) \leq k (t-t_{1})
\end{equation}
$\forall t> t_{1}$ (note that $f(t,s))>0$), and $\forall s \in [0,b]$. But using Cauchy-Schwarz inequality in (\ref{4.3a}):
\begin{equation}
\label{4.3b}
f(t,s) \leq \frac{\parallel \delta \parallel + \sqrt{\parallel \delta \parallel ^{2} + \beta \cdot \lambda _{max}}}{\beta}(t(s),x(s))
\end{equation}
Thus, under our hypothesis, putting $K= {\rm Max} \{ K(x(s)): s\in [0,b]\} $, we have:
\[
f(t,s)  \leq (1+ \sqrt{2}) K t  
\]
for $t>1$, and the condition (\ref{4.3 cielos}) can be achieved. $\Box$

\vspace{3mm}

\noindent {\it Remark}. (1) It may be more convenient (even though it is not more general), to replace the bound $K(x)t$ in Proposition \ref{prop 4.3} by $K_{1} t + K_{2}$, where $K_{1}$ and $K_{2}$ are two functions on ${\cal M}_{0}$ (or even to put different functions for each inequality).  

(2) The continuity of $K$ is necessary. So, the bounds in (\ref{bound}) can be restated as: for each $x_{0} \in {\cal M}_{0}$ there exist a neighborhood $U \ni x_{0}$ such that $\parallel \delta \parallel /(\beta t)$ and 
$\lambda_{max}/(\beta t^{2})$ are bounded in $U\times ]1,\infty[$. 
Note that the behaviour of the metric for $t\rightarrow -\infty$ is not relevant (for the finiteness of $T_{0}^{*}$,  it is not  the behaviour for $t\rightarrow \infty$). 

\vspace{3mm}

\noindent {\it Example.} The pseudosphere $S^{n}_{1}$ (de Sitter spacetime) can be written as a product $\R{}\times S^{n-1}$ with $(S^{n-1}, <\cdot , \cdot>)$ the canonic Riemannian $(n-1)$-sphere, and a metric $g$ with $\beta \equiv 1$, $\delta \equiv 0$, $\alpha (t,x)= cosh^{2}(t) \cdot$Id; by Corollary \ref{coro 3.3} it is globally hyperbolic. The growth of $\lambda_{max} /\beta$ in this case is exponential, and $T_{0}$  reaches the value $\infty$. In fact, a straightforward computation shows $T_{0}((t,x),-x) = \infty$ for all $(t,x)$ (see \cite[Proposition 5.38]{On}). 
More precisely, consider a globally hyperbolic warped metric $(\R{} \times {\cal M}_{0} , g=-dt^{2} + f^{2}(t) <\cdot , \cdot >)$ as in Example {\it (2.B)} at the end of Section 3. In this case, equation 
(\ref{4.3a})  becomes $t'(s) = f(t(s))$ and, thus, 
\[
T_{0}((t_{1},x_{1}),x_{2}) < \infty \; \; \,{\rm if\; and\; only\; if\; } \; \, \int_{t_{1}}^{\infty} \frac{dt}{f(t)} > {\it dist}(x_{1},x_{2}) \, .
\]
So, $T_{0}$ remains finite on all ${\cal M \times M}_{0}$ if and only if 
$\int_{0}^{\infty} dt/f(t) = \infty $. 
 $\Box$

\vspace{3mm}

\noindent As a consequence of Proposition \ref{prop 4.3}, we have:

\begin{coro}
\label{coro 4.4}
Assume that each slice ${\cal M}_{t}$ of ${\cal M}$ is a Cauchy surface, and $\alpha , \beta,  \delta $ have a periodic dependence of period $\Delta$ with the universal time $t$ (in particular, when the spacetime is stationary). Then:

(i) $T_{0}$ is never infinite, and

(ii) The quotient ${\cal M}/\Delta$ obtained by identifying each pair $(t,x), (t+\Delta ,x) \in {\cal M}$ is geodesically connected by timelike geodesics (in particular, it is totally vicious). 
\end{coro}

{\it Proof.} The first assertion is obvious from the Proposition \ref{prop 4.3}. For the second, take any $z_{1} = (t_{1}+\Delta \Z{} , x_{1}), z_{2} = (t_{2}+\Delta \Z{} ,x_{2}) \in {\cal M}/\Delta$ and choose $n\in \N{}$ such that $t_{2}+n\Delta > T_{0}((t_{1}, x_{1}),x_{2})$. Then $(t_{1}, x_{1})$ and $(t_{2}+n\Delta , x_{2})$ can be joined by a timelike geodesic, which projects in the required one. On the other hand, any spacetime such that for each couple of points there exists a timelike curve joining them is clearly totally vicious (see, for example, \cite{Mat}). $\Box$
 
\vspace{3mm}

\noindent When the function $T_{0}$ is finite, Avez-Seifert result can be used to yield the next multiplicity result. Given $(t_{1}, x_{1}), (t_{2},x_{2})\in {\cal M}$ call $N((t_{1}, x_{1}), (t_{2},x_{2}))$ to the number of future-directed timelike geodesics joining the first point with the second one (up to reparametrizations).

\begin{prop}
\label{prop 4.5}
Assume that each slice ${\cal M}_{t}$ of ${\cal M}$ is a Cauchy surface, and the inequalities {\rm (\ref{bound})} hold. Let $p$ be the cardinal of the fundamental group $\pi ({\cal M}_{0})$, 
and fix $(t_{1},x_{1}) \in {\cal M}$ :

(i) If $p$ is finite there exists $\Delta >0$ such that 

\vspace{-1mm}
\[
N((t_{1}, x_{1}), (t_{2},x_{2})) \geq p
\]

\vspace{-1.5mm}
\noindent for all $t_{2} > t_{1} + \Delta$.

(ii) If $p$ is infinite then 
\[
\lim_{t_{2} \rightarrow \infty} N((t_{1}, x_{1}), (t_{2},x_{2})) = \infty .
\]
\end{prop}

{\it Proof.} Consider the universal Lorentzian covering $\pi: {\cal M}'={\cal M}'_{0}\times \R{} \rightarrow {\cal M}$. The slices of this spacetime are also Cauchy surfaces and it is straightforward  to check from Proposition \ref{prop 4.3} that the corresponding arrival function $T_{0}'$ is finite. Then, the result follows fixing 
$(t_{1}', x_{1}') \in \pi ^{-1} (t_{1}, x_{1})$, connecting it by a timelike geodesic with  the static trajectories at each point of $\pi ^{-1} (t_{2}, x_{2})$, and projecting these geodesics in ${\cal M}$. $\Box$

\vspace{3mm}

\noindent Note that when $T_{0}((t_{1},x_{1}),x_{2}) \geq t_{2}-t_{1} > 0 $   then $(t_{1},x_{1})$ and $(t_{2},x_{2})$ are not chronologically related (Lema \ref{lema 4.1}) and, in particular, $N((t_{1},x_{1}),(t_{2},x_{2}))=0$. Then, as a straightforward consequence:

\begin{lema}

\label{lema 4.6}
Given  $(t_{1},x_{1}) \in {\cal M}$ and $x_{2} \in {\cal M}_{0}$, $x_{2} \not= x_{1}$, there exist $\Delta >0$ such that:
$t-t_{1}\leq \Delta$ if and only if   $N((t_{1},x_{1}),(t,x_{2}))=0$.
\end{lema}

{\it Proof.}  Take $\Delta =  T_{0}((t_{1},x_{1}),x_{2})$, and note that $T_{0}$ vanishes only on the couples $((t,x),x)\in {\cal M}\times {\cal M}_{0}$. $\Box$

\vspace{3mm}

\noindent {\it Remark.} Taking $\Delta ^{*} = T_{0}^{*}((t_{1},x_{1}),x_{2})$ the next property holds:
{\it $t-t_{1}\geq -\Delta ^{*}$ if and only if $N((t,x_{2}),(t_{1},x_{1})) = 0$. }

So, if $T_{0}=T_{0}^{*}$ (in particular, in the stationary case)  the constant $\Delta  = T_{0}((t_{1},x_{1}),x_{2})$ satisfies: {\it $\mid t-t_{1}\mid \leq \Delta $ if and only if $(t,x_{2})$ and $(t_{1},x_{1})$ are not chronologically related} (if and only if  $N((t,x_{2}),(t_{1},x_{1})) = N((t_{1},x_{1}),(t,x_{2})) = 0$). 

\vspace{3mm}

\noindent On the other hand, from (\ref{4.3a}) it is not difficult to impose conditions yielding uniform bounds from below to $\mid t' \mid$, obtaining so obstructions to the chronological relations and sufficient conditions for $N$ to vanish, as in the next result.

\begin{prop}
\label{prop 4.7}
Assume that:
\begin{equation}
\label{emes}
 \parallel \delta \parallel \leq M_{\delta},\;\;  m_{\alpha } \leq \lambda_{min} , \;\; m_{\beta }<\beta <M_{\beta}
\end{equation}
 for some $M_{\delta }, m_{\alpha }, m_{\beta }, M_{\beta} >0$. Then there exist $m>0$ such that: 
if $\mid t_{2}-t_{1} \mid \leq m \cdot {\it dist}(x_{1}, x_{2}) $ then $(t_{1},x_{1})$ and $(t_{2},x_{2})$ are not chronologically related 
(or, equivalently,   $N((t_{1},x_{1}),(t_{2},x_{2})) = N((t_{2},x_{2}),(t_{1},x_{1})) = 0$). 
\end{prop}

\noindent (Note that the inequalities (\ref{emes}) imply that each slice ${\cal M}_{t}$ is a Cauchy surface by Proposition \ref{prop 3.2}). 

\vspace{3mm}
 
{\it Proof.} Consider any timelike curve $\gamma = (t,x), \parallel x' \parallel = 1$ joining two given points  $(t_{1},x_{1}),(t_{2},x_{2})$. From the inequality corresponding to (\ref{4.3a}):
\[
\mid t'(s) \mid > -\frac{\mid <\delta ,x'> \mid}{\beta} (t(s),x(s)) + \sqrt{\frac{<\delta,x'>^{2}}{\beta ^{2}} + \frac {<\alpha (x'),x'>}{\beta}}(t(s),x(s)) 		
\] 
\[
\geq - \frac{M_{\delta}}{m_{\beta}} + \sqrt{\frac{M_{\delta} ^{2}}{m_{\beta} ^{2}} + \frac{m_{\alpha}}{M_{\beta}}} \equiv m
\]
Integrating it 
\[
\mid t_{2} - t_{1} \mid > m \cdot {\rm length}(x) \geq m \cdot {\it dist}(x_{1},x_{2}) 
\]
from which the result follows.
$\Box$

\vspace{3mm}

\noindent Note that the greater $m$ above is:
\[
{\rm Inf}\{ \frac{T_{0}((t,x_{1}),x_{2})}{{\it dist}(x_{1},x_{2})}, \frac{T_{0}^{*}((t,x_{1}),x_{2})}{{\it dist}(x_{1},x_{2})} : (t,x_{1})\in {\cal M} , 
x_{2} \in {\cal M}_{0} , \; {\rm and} \; x_{2} \not= x_{1} \}
\]
Lemma \ref{lema 4.6} and Proposition \ref{prop 4.7} should be 
compared with the analogous ones obtained from VM (for example 
\cite[Theorem 3.5.2]{Ma-li} , see also \cite{Sa-ca}).

\section{Open questions and comments.}
Now, we can compare the results in the last two sections with those obtained by using VM. 

\begin{enumerate}
\item The hypothesis in Propositions \ref{prop 3.1} or \ref{prop 3.2} for global hyperbolicity are less strong than those in \cite[Theorem 1.1]{GM-ihp} for the existence of a timelike geodesic. On the other hand, under these stronger assumptions there it is obtained the next variational characterization of the chronological relation:
two points $z_{1}=(t_{1},x_{1}), z_{2}=(t_{2},x_{2}) \in {\cal M}$ are chronologically related if and only if 
\begin{equation}
\label{5.1}
{\rm Sup}_{t\in C^{1}(t_{1},t_{2})} {\rm Inf}_{x\in C^{1}(x_{1},x_{2})} \int _{0}^{1} \{ <\alpha _{z}(x'),x'> + 2<\delta (z),x'> -\beta(z)t'^{2}\}ds < 0 
\end{equation}
where $C^{1}(t_{1},t_{2})$ (resp. 
$C^{1}(x_{1},x_{2})$) denotes the set of $C^{1}$-curves defined on $[0,1]$ which join $t_{1}$ and $t_{2}$ (resp. $x_{1}$ and $x_{2}$), and 
$z=(t,x):[0,1]\rightarrow {\cal M}$.
This characterization (\ref{5.1}) is analogous to the one in \cite{Gi}, \cite{GM-man} for the existence of a timelike geodesic between two points in manifolds as the second  region of Reissner-Nordstr\"{o}m spacetime. In \cite{Sa-grg} it is shown that (\ref{5.1}) can be also interpreted  as a characterization of the chronological relation  for these spacetimes. A related characterization for the stationary case is obtained in \cite[Lemma 3.5.1]{Ma-li} under assumptions which imply global hyperbolicity. Then, it is natural to wonder: {\it at what extent is the relation {\rm (\ref{5.1}) } valid as a variational characterization of the chronological relation?}

\item Global hyperbolicity and the finiteness of the functions $T_{0}, T_{0}^{*}$ are usually imposed from a VM point of view to obtain geodesic connectedness. For example, the conditions $(i), (ii)$ in \cite[Theorem 3.4.3]{Ma-li} or (1.2), (1.3), (1.5) and (1.6) in \cite{GM-ihp} imply global hyperbolicity (compare them with Corollary \ref{coro 3.4} and Proposition \ref{prop 3.2}, resp.) On the other hand, 
the conditions in Proposition \ref{prop 4.3} for the finiteness of $T_{0}$, and the analogous ones for the finiteness of $T_{0}^{*}$, are slightly weaker than the conditions (1.9) in \cite{GM-ihp} (see also \cite[Theorem 1.1]{BFM}). In this reference \cite{GM-ihp}, this condition is introduced to make sure the topological non-triviality of the sublevels of the energy functional. Note also that stationary spacetimes automatically satisfy the finiteness of $T_{0}$ and $T_{0}^{*}$, and so, this condition is normally imposed from the VM point of view to obtain results on geodesic connectedeness. In fact, anti-de Sitter and de Sitter spacetimes  
are counterexamples showing that neither global hyperbolicity nor the finiteness of $T_{0}$ and $T_{0}^{*}$ can be removed to obtain geodesic connectedness. 
Nevertheless,  some technical additional conditions are assumed to obtain it from the VM point of view (see, for example, \cite{GM-ihp}) and it is natural to wonder at what extent are these technical assumptions necessary. That is, 

\hspace{5mm} {\it if each slice ${\cal M}_{t}$ of $({\cal M},g)$ is a Cauchy surface and the functions $T_{0}, T_{0}^{*}$ remain finite, is $({\cal M}, g)$ geodesically connected? }

\item Consider the multiplicity result in Proposition \ref{prop 4.5} {\it (ii)} joined with 
the Propositions \ref{prop 3.1} or \ref{prop 3.2} on global hyperbolicity.  These results together can be seen as a multiplicity result for the existence of timelike geodesics with hypothesis:
\begin{itemize}

\item weaker for the coefficients $\alpha, \beta, \delta$ than those in \cite[Theorem 1.4]{GM-ihp}, \cite[Theorem 3.5.6]{Ma-li}, but 

\item stronger for the topology of the manifold ${\cal M}_{0}$ than those in the quoted references (there it is assumed that ${\cal M}_{0}$ is not contractible, no assumption on $\pi ({\cal M}_{0})$ is done). 
\end{itemize}

Then it is natural to  wonder: {\it is it possible to improve {\rm Proposition \ref{prop 4.5}{\it (ii)}} by imposing to ${\cal M}_{0} \,$ just to be contractible?}

\end{enumerate}

\section*{Acknowledgments}
The main part of this work has been developed during a two-months stay of the author at the University of Bari supported by a EC Contract (Human Capital and Mobility) ERBCHRXCT940494. The author wants to acknowledge 
to the members of the Dipartimento di Matematica of this University for their kind hospitality. Very especially, he is grateful to Prof. D. Fortunato for his attention, discussions  and suggestions. On the other hand, the author acknowledges warmly to Prof. P.E. Ehrlich his (electronic) comments on this paper.

\end{document}